\documentclass{article}

\usepackage{fancyhdr}
\usepackage{extramarks}
\usepackage{amsmath}
\usepackage{amsthm}
\usepackage{amsfonts}
\usepackage{tikz}
\usepackage[plain]{algorithm}
\usepackage{algpseudocode}
\usepackage{enumerate}
\usepackage{tikz}
\usepackage{amssymb}
\usepackage{bm}
\usepackage{indentfirst} 
\usepackage{multicol}
\usepackage{hyperref}
\usepackage{graphicx}
\usepackage{float}
\usepackage{array}

\hypersetup{hidelinks,
	colorlinks=true,
	allcolors=blue,
	pdfstartview=Fit,
	breaklinks=true}

\newcommand{\thesisTitle}{Improvements on Recommender System based on Mathematical Principles}

\newcommand{\thesisAuthorName}{Authors: Fu Chen, Junkang Zou, Lingfeng Zhou, Zekai Xu, Zhenyu Wu}

\title{
    \vspace{2in}
    \textmd{\textbf{\thesisTitle}}\\
   \vspace{2in}
}

\author{
             \text{\thesisAuthorName} 
        }
\date{}

\begin{document}

\maketitle
\thispagestyle{empty}
\pagebreak

\begin{abstract}
    In this article, we will research the Recommender System's implementation
 about how it works and the algorithms used. We will explain the Recommender System's
  algorithms based on mathematical principles, and find feasible methods for improvements.
   The algorithms based on probability have its significance in Recommender System, we will describe
    how they help to increase the accuracy and speed of the algorithms. Both the weakness and the strength
     of two different mathematical distance used to describe the similarity will be detailed illustrated in this article.
\end{abstract}
\thispagestyle{empty}

\tableofcontents

\thispagestyle{empty}

\pagebreak

\begin{multicols}{2}

\setcounter{page}{1}

\section{Introduction}
    Generally, Recommender System is used to sift out desirable messages from a
 large amount of data based on some algorithms. The algorithms like algorithm applied for ranking,
 the algorithm applied for filtering are the core of a Recommender System. If we ignore the algorithms'
 mathematical principles, the speed and the accuracy can be unpromising. \\
  \par In order to group users that have similar
 preference to improve efficiency, we need to quantify the degree of similarity between them according to mathematical distance and similarity function. 
 After several determinant aspects are chosen, user's preference can be simplified into a vector. Based on the vector, the comparison
 of the variety of algorithms can be conducted. The idea above leads to this short article: algorithms based on Mathematical statistics not
 only help to accurately filter out large amount of useless data but also decrease the time used to calculate in the ranking process.

\section{Implementation of Recommender System}
    In this chapter, we are going to introduce the most basic Recommender System Model.\\
    \par The Recommender System generally consists of two modules: \textbf{Offline Algorithm Module} and \textbf{Real-time Algorithm Module}.\\
    \par The \textbf{Offline Algorithm Module} trains several models including ranking model based on historical data and generates offline preference result. The Offline Module typically updates 
            the data everyday. Since it has 24 hours to compute the preference, the concrete algorithms in this module can be complicated.\\
    \par The \textbf{Real-time Algorithm Module} collects latest data and then modifies it with the feedback from the Offline Algorithm Module.\\
    \par Below is an example of the implementation of Recommender System in video application:
        \begin{center}
        $\includegraphics[scale=0.4]{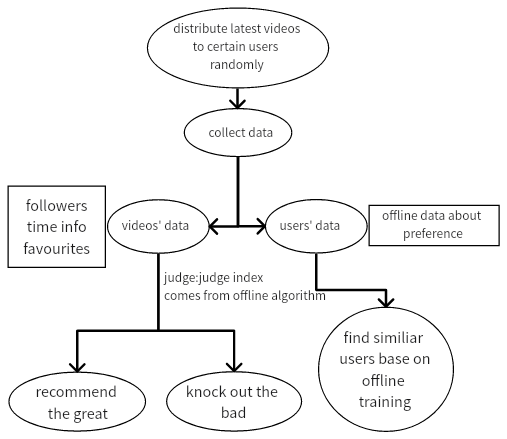}$
        \end{center}
    \par In the rest of this article, we will focus on the \textbf{Real-time Algorithm Module}, introducing its basic algorithm and giving potentially feasible optimization methods.

\section{Real-time Algorithm and improvements based on Mathematical Principles}
    \subsection{Collaborative Filtering Algorithm}
    Collaborative Filtering Algorithm is the most widely used algorithm in Real-time Algorithm Module. It first divides users into certain groups
 by their historical preference similarity and then gives latest recommendation
 based on other group members' feedback.
        \subsubsection{Implementation Process}
            Firstly, we determine several aspects that is enough to describe users' preference by assumption. Then we collect users' data of predetermined aspects to form a vector for each user.\\
            \par Secondly, we choose an appropriate mathematical distance to measure the similarity of users'
         preference. Below are two commonly used mathematical distances:\\
        \\  1.\textbf{Euclidean Distance}
            \begin{equation*}
                d_{E}(\textbf{A},\textbf{B})=\sqrt{(a_1-b_1)^2+...+(a_n+b_n)^2}
            \end{equation*}
            where A=($a_1,...,a_n$), B=($b_1,..,b_n$)\\
        \\  2.\textbf{Mahalanobis Distance}
            \begin{equation*}
                d_{M}(\textbf{A},\textbf{B})=\sqrt{(\textbf{A}-\textbf{B})^{T}\bm{S^{-1}}(\textbf{A}-\textbf{B})}
            \end{equation*}
            where $\bm{S^{-1}}$ is the sample covariance matrix.\\
            \par Then we use the \textbf{similarity function} to measure the similarity of different users:
            \begin{equation*}
                sim(\textbf{A},\textbf{B})=\frac{1}{1+d(\textbf{A},\textbf{B})}
            \end{equation*}
            \par Lastly, we give preference similarity based on the similarity distance between users. The larger value the sim() function returns, the more similar the users' preference. 
        \subsubsection{Weakness and Strength of two mathematical distance}
            The Euclidean Distance is easier to compute, since it only involves the operation of subtraction, 
            addition, square and square root. \\
            \par However, since Euclidean Distance equates the importance of all components, the less significant component might deviate the measured similarity from the true one. To fix this incommensurate value problem, we need to look no further than the standardization process.\\ 
            \par Moreover, the result might be influenced by the component that has dominant absolute value. For example, 10 mm height gap and 10cm height gap are totally different in percentage change of height but they are viewed as the same in Euclidean distance.\\
            \par The Mahalanobis Distance uses the sample covariance matrix to minimize the correlation between different components and assigns weight to different components to indicate its importance. \\
            \par However, since the inverse of sample covariance matrix is hard to calculate and may not exist, we shall be careful when using Mahalanobis Distance.\\
    
    \subsection{Optimization of Filtering Algorithm}
        Though Collaborative Filtering Algorithm can reflect users' preference comprehensively, its processing speed is always unsatisfactory, which inspires us to optimize it with mathematical principles.\\
        \par In this section, we are going to introduce two mathematical statistics' applications in optimizing the algorithm.
    \subsubsection{Filtering Algorithm with Multivariate Normal Prior Distribution}
            With Multivariate Normal Prior Distribution, the sample (long-term historical data) also obeys multivariate normal distribution. And since multivariate normal distribution
        can be determined by its first two order moments, all the thing that the algorithm has to do is to estimate parameters.\\
        \par The PDF of estimated sample distribution is given below:
        \begin{equation*}
            f_{\textbf{X}}(x_1,...,x_m)=\frac{1}{(2\pi)^{\frac{m}{2}}|\textbf{S}|^{\frac{1}{2}}}e^{-\frac{1}{2}(\bm{X-\mu})^T\bm{S^{-1}}(\bm{X-\mu})}
        \end{equation*}
        \par Given the predetermined determinant index between 0 and 1 and the estimated parameters, we compare the determinant index with the value of CDF of the multivariate normal distribution and assign corresponding weight to components to indicate its importance.

    \subsubsection{Filtering Algorithm with Quantile Estimation}
        The Multivariate Normal Prior Distribution is too strict for most cases. A more widely used method is not to have a prior assumption and use the \textbf{Quantile Estimation}. To make formulas simple, we only consider unidimensional situation.\\
        \par We first give some theoretical foundations about Quantile Estimation:\\
        \\ (1)\textbf{p-Quantile Point} $\xi_p$
        \begin{equation*}\begin{aligned}
            Prob(X<\xi_p) &\leq p\\
            Prob(X\geq \xi_p) &\leq 1-p
        \end{aligned}\end{equation*}
        (2)\textbf{Order Statistics of Sample}\\
        \\ If we have sample $X_1,...,X_n$ that is independent identically distributed(i.i.d.) and we sort it into increasing order and get Order Statistics $X_{(1)}\leq ...\leq X_{(n)}$\\
        \\ (3)\textbf{Test Statistics}\\
        \\ Naturally, $X_{[np]}$ is an estimator of $\xi_p$\\
        \par To meet with certain significance level needs, we could construct a confidence interval using the following result:
        \begin{equation*}
            P(X_{(r)}\leq \xi_p < X_{(s)})=\sum_{i=r}^{s-1} \left(\begin{matrix}
                                                                        n\\
                                                                        i\\
            \end{matrix}\right)p^i(1-p)^{n-i}
        \end{equation*}
        \indent Obviously, this method doesn't require any distribution assumptions. However, the exactness could be undesirable compared with method with prior assumption. The concrete code is given in the paper given by Keith in \textbf{Algorithm1}. The proof of the formula above is also in the pape. \\
        \par Besides, the Quantile Estimation could also be used in the \textbf{Offline Algorithm Module} to modify and verify results of other models.

    \subsection{BM25 Algorithm}
        In this section, we are going to introduce BM25 Algorithm used in search engines and try to apply it in the Recommender Algorithm System.
        \subsubsection{Corpus and Sentence Division}
            We first introduce the \textbf{Corpus and Sentence Division}, foundation of \textbf{BM25 Algorithm}.\\
            \par Corpus means the collection of text and Sentence Division means to divide sentence according to certain rules. Corpus and Sentence Division is important for search engines. 
                If the sentence is divided into too small parts, the number of individuals with the same meaning will be larger, and personalized degree may not be guaranteed. An very commonly used standard is to categorize
                by synonym, antonym and idiom.\\
            \par \textbf{TF-IDF(Term Frequency-Inverse Document Frequency)} is often used to extract keywords from articles. It can be defined as the calculation of how relevant a word in a series or corpus is to a text.
             The meaning increases proportionally to the number of times in the text a word appears but is compensated by the word frequency in the corpus. TF represents the frequency of a Term in the Document, and the higher the frequency, the more important it is.
              DF represents the total number of Documents that contain this word. The larger the DF, the more common the word, and the less important it is. The smaller the DF, the more important it is. And IDF is a function of DF, so that the larger the IDF, the more important the word. \\
            \par If the frequency relationship between words in the article is that "Blink" > "practice" > "conclusion" then we can say that "Blink" is the most important keyword for this article. However, if we find that the three words
             "Blink," "practice" and "conclusion" have the same frequency, it does not mean that they are equally important as keywords. It is because the total number of blogs containing these keywords is "Blink" < "practice" < "conclusion"
             , indicating that "Blink" is not that common. But when it appears, it is more important than "practice" and "conclusion" for this article. 

        \subsubsection{BM25 Algorithm}
            Based on the information given in Corpus and Sentence Division section, we are going to introduce the BM25 Algorithm. We first give its definition:
            \begin{small}\begin{equation*}\begin{aligned}
                score(D,Q) & =\sum^{n}_{i=1}IDF(q_i) \frac{(k_1+1)f(q_i,D)}{f(q_i,D)+k_1(1-b+b\frac{|D|}{avgdl})}\\
                IDF(q_i) & =ln(\frac{N-n(q_i)+0.5}{n(q_i)+0.5}+1)\\
            \end{aligned}\end{equation*}\end{small}
            \noindent $\bm{q_i}$\, denotes a certain word or phrase\\
            \noindent $\bm{f(q_i,D)}$\, denotes the time that label $q_i$\, appears in file $D$\\
            \noindent $\bm{|D|}$\, denotes the length of document $D$\, in the text collection from which documents are drawn\\
            \noindent $\bm{k_1}$\, and $\bm{b}$\, are free parameters based on the results of \textbf{Offline Algorithm Module}\\
            \noindent \textbf{Avgdl} is the average length \\
            \noindent \textbf{Q} and \textbf{D} stand for query and document respectively\\
            \par   \textbf{Score(Q, D)} describes the matching degree of D and Q based on words and phrases Though the \textbf{BM Algorithm} is used in search engines, it has reference significance
             in label recommendation in Recommender System. For example, IDF can be used to calculate the weight of label, the number of synonyms in the video when giving videos recommendation. 

        \subsubsection{Optimization based on BM25 Algorithm}
            In this section, we are going to give a potentially feasible optimization method based on \textbf{BM Algorithm} in \textbf{Real-time Algorithm Module}.\\
            \par Firstly, we shall divide all labels into appropriately small labels and all the synonymous labels should be classified into one group.\\
            \par Secondly, if we treat all the labels from one object a query in the BM25 Algorithm and all the labels in all the objects a document, and naturally we could apply the BM25 Algorithm in Recommender System. \\
            \par But as mentioned before, we must pay attention to the similarity and importance of certain labels, whose information generally lies in the IDF.
            Typically, video or other objects' tags don't repeat as much as searching for documents. So the BM25's function is enough to satisfy our needs, if given some adjustment. So it spires us to devise the score formula based on specific needs. We couldn't
             give the detailed formula, but only illustrate the potential of this algorithm in Real-time Algorithm Module.\\
            \par As for the judgement of the score value, we shall look back to the \textbf{Quantile Estimation} in the previous chapter. If the value is 
            larger than value of Quantile Estimation whose standard is predetermined, then the object should be recommended.

\section{Conclusion}
    In this article, We analyze the Recommender System in detail, and its related mathematical principles in this article.\\
    \par In the first part of this article, we first introduce the two modules the common Recommender System consists of and how these two modules interact with each other. \\
    \par Then we dig deep in the \textbf{Real-time Algorithm Module}. We begin with \textbf{Filtering Algorithm} and its two different mathematical distances, illustrating both its strength and weakness. And we 
            give two possible methods to improve the algorithm based on mathematical principles.\\
    \par In the last part, we try to take a page from search engines' book and apply \textbf{BM25 Algorithm} to Recommender System. We have listed basic concepts involved in search engines' algorithm and given the 
    score function used in BM25 Algorithm. Based on this, we come up with the several possible modification of this function to meet needs of Recommender System.

\section{Acknowledgement}
    Keith Briggs and Fabian Yang introduce the detailed technique and algorithm of Quantile
     Estimation. Its practical significance is revealed in our improvements on Recommender System.

\section{Reference}
    \begingroup 
    \renewcommand{\section}[2]{} 
    
    \endgroup

\end{multicols}

\end{document}